\documentclass[manuscript]{aastex6}

\usepackage{CJK}

\shorttitle{Discovery of A New Retrograde Trans-Neptunian Object}
\shortauthors{Ying-Tung Chen et al. }

\begin{document}
\begin{CJK*}{UTF8}{bkai}


\title{Discovery of A New Retrograde Trans-Neptunian Object:\\ Hint of A Common Orbital Plane for Low Semi-Major Axis, High Inclination TNOs and Centaurs}




\author{Ying-Tung Chen (陳英同)}
\affil{Institute of Astronomy and Astrophysics, Academia Sinica,
P. O. Box 23-141, Taipei 106, Taiwan}
\email{ytchen@asiaa.sinica.edu.tw}
\author{Hsing Wen Lin (林省文)}
\affil{Institute of Astronomy, National Central University, 32001, Taiwan}
\author{Matthew J. Holman, Matthew J. Payne}
\affil{Harvard-Smithsonian Center for Astrophysics, 60 Garden Street, Cambridge, MA 02138, USA}
\author{Wesley C. Fraser}
\affil{Queen's University Belfast, Astrophysics Research Centre, Belfast, UK}
\author{Pedro Lacerda}
\affil{Max-Planck-Institut f\"ur Sonnensystemforschung, Justus-von-Liebig-Weg 3, D-37077 G\"ottingen, Germany}
\author{Wing-Huen Ip (葉永烜)}
\affil{Institute of Astronomy, National Central University, 32001, Taiwan}
\affil{Space Science Institute, Macau University of Science and Technology, Macau}
\author{Wen-Ping Chen (陳文屏)}
\affil{Institute of Astronomy, National Central University, 32001, Taiwan}
\author{Rolf-Peter Kudritzki, Robert Jedicke, Richard J. Wainscoat, John L. Tonry, Eugene A. Magnier, Christopher Waters, Nick Kaiser}
\affil{Institute for Astronomy, University of Hawaii at Manoa, Honolulu, HI 96822, USA}
\author{Shiang-Yu Wang, Matthew Lehner}
\affil{Institute of Astronomy and Astrophysics, Academia Sinica, P. O. Box 23-141, Taipei 106, Taiwan}
\begin{abstract}
Although the majority of Centaurs are thought to have originated in the scattered disk, with the high-inclination members coming from the Oort cloud, the origin of the high inclination component of trans-Neptunian objects (TNOs) remains uncertain.
We report the discovery of a retrograde TNO, which we nickname ``Niku'', detected by the Pan-STARRS 1 Outer Solar System Survey.  Our numerical integrations show that the orbital dynamics of Niku are very similar to that of 2008 KV$_{42}$ (Drac), with a half-life of $\sim\,500$\,Myr. Comparing similar high inclination TNOs and Centaurs ($q > 10$\,AU, $a < 100$\,AU and $i > 60\arcdeg$), we find that these objects exhibit a surprising clustering of ascending node, and occupy 
a common orbital plane. 
This orbital configuration has high statistical significance: 3.8-$\sigma$.
An unknown mechanism is required to explain the observed clustering. This discovery may provide a pathway to investigate a possible reservoir of high-inclination objects.
\end{abstract}

\section{Introduction} \label{sec:intro}
Many primitive bodies exist in the vast regions of the solar system beyond Jupiter, of which the largest population is the trans-Neptunian objects (TNOs). The details of the orbital distribution of the TNOs preserve information about the evolution of the solar system \citep{Levison03, Lykawka07}. Following the evolution of the planetesimal disk, most TNOs were left in low inclination orbits
\citep{Levison08}. Even scattered disk objects 
have typical inclinations less than 30$\arcdeg$-40$\arcdeg$ \citep{Gomes05}. However, the discovery of 2008 KV$_{42}$ \citep{Gladman09} revealed the first member of a new population: a retrograde TNO. Dynamical simulations of 2008 KV$_{42}$ demonstrate that it has a very long lifetime (a few Gyr) \citep{Gladman09}, suggesting that a large population with similar orbits may exist in this region.

Centaurs are minor planets with semi-major axes between those of Jupiter and Neptune, and whose orbits are planet-crossing. They constitute a link between the short-period Jupiter-family comets and the reservoir of icy bodies in the outer Solar System \citep{Levison97}. 
Centaur orbits are typically unstable, with 
lifetimes of $5\sim50$\,Myr \citep{Volk13}. Centaurs are generally assumed to originate as scattered disk objects which interact with Neptune, causing a change in semi-major axis. However, high inclination Centaurs seem not to originate from the scattered disk (whose inclination distribution seems to be too narrow 
to supply high inclination objects such as 2008 KV$_{42}$). \citet{Volk13} also find that the Kuiper belt is an extremely unlikely source of the retrograde Centaur. Some researchers suggest that the Oort cloud could be the source of such high inclination Centaurs \citep{Brasser12, Rabinowitz13}, but given the small number of high inclination objects in the Minor Planet Center (MPC) catalogs, the origin of the high inclination population in the outer Solar System will remain uncertain until more such objects are detected and their orbital distribution is understood.

The Panoramic Survey Telescope and Rapid Response System 1 Survey (Pan-STARRS 1, hereafter PS1) is the first wide-field optical system ($3\pi$ steradians) using a dedicated large aperture (1.8 m) telescope to carry out multi-epoch, multi-color observations with careful calibration  \citep{Tonry12,Schlafly12,Magnier13}. 
The PS1 Outer Solar System (OSS) key project has completed an initial search for slow-moving objects, resulting in hundreds of candidates, of which $\sim\,50\%$ are known \citep{Holman15}. 
In the catalog of initial results, we identified a distant object with an inclination greater than $90\arcdeg$, i.e. retrograde TNO.

Here we report the discovery of a second retrograde TNO and analyze its dynamical evolution.   In addition, after selecting known objects from the MPC database which satisfy the orbital criteria of perpendicular orbits
, with the aim of identifying similar objects, we found a population of high-inclination objects that occupy the same orbital plane but which orbit in both senses (retrograde and prograde).   

\section{Observations} \label{sec:obs}
The PS1 survey observed the entire sky north of declination $-30\arcdeg$ using a Sloan-like filter system ($g_{P1}$, $r_{P1}$, $i_{P1}$, $z_{P1}$, $y_{P1}$).  In addition, PS1 observed within $\pm20\arcdeg$ of the plane of the ecliptic using the $w_{P1}$-band (limiting magnitude $\sim22.5$, correspond to TNO $H \sim 6.5$), which spans the wavelength range of $g_{P1}$, $r_{P1}$ and $i_{P1}$. 
Upon searching the PS1 for outer solar system objects, we identified an unusual object, which we nicknamed “Niku”, with a retrograde, nearly polar orbit.

Niku has been observed 22 times by PS1 at two different oppositions.  To test and improve the orbit determination of Niku, we obtained follow-up observations with the 1-m Lulin Observatory Telescope (LOT) in Taiwan. In addition, we gathered archival DECam and CFHT observations using SSOIS (Solar System Object Image Search\footnote{\url{http://www.cadc-ccda.hia- iha.nrc-cnrc.gc.ca/en/ssois/}}) \citep{Gwyn12}. The astrometry and photometry of all observations were calibrated against the PS1 catalog.  We determined the orbit of Niku, based on observations spanning four oppositions, using the orbit fitting code of \citep{BK00}. The resulting uncertainties in the heliocentric orbital elements are all small. 
The elements are inclination $i = 110.2929\pm0.0004 \arcdeg$, longitude of ascending node $\Omega = 243.8181\pm0.0001 \arcdeg$, semi-major axis $a = 35.724932\pm0.006153$\,AU, eccentricity $e = 0.333599\pm0.000144$; and argument of pericenter $\omega = 322.593\pm0.015$.  Pericenter passage will occur at $2451287.350\pm1.105$. The current barycentric distance of Niku is $25.892\pm0.001$, well inside the orbit of Neptune.  We independently verified the orbit determination using the OpenOrb package of \citep{Granvik09}; the results are the same within uncertainty. We note that the orbital elements of Niku are very similar to those of the  first retrograde TNO, 2008 KV$_{42}$. Their semi-major axes are both beyond Neptune's orbit; eccentricities are in the range $0.3\sim0.5$ and the inclinations of both are larger than $100\arcdeg$

After we submitted the astrometry of Niku to the MPC, Niku was linked with 2011 KT$_{19}$, an object with a short observational arc (8 days) The initial orbit of 2011 KT$_{19}$ was identified as a prograde Centaur (MPEC 2011-L09, $a = 27.6, e = 0.41, i = 38.02$). The combination of the MPC data for 2011 KT$_{19}$ and our observations somewhat improves the orbit determination.   However, our observations of Niku alone are good enough to perform a reliable dynamical analysis.  Considering the internal consistency of the PS1 reference frame,  we use the orbit based on the data measured with the PS1 star catalog for all further analyses.

\section{Numerical Integrations and Analysis} \label{sec:analysis}
\begin{figure}
\figurenum{1}
\plotone{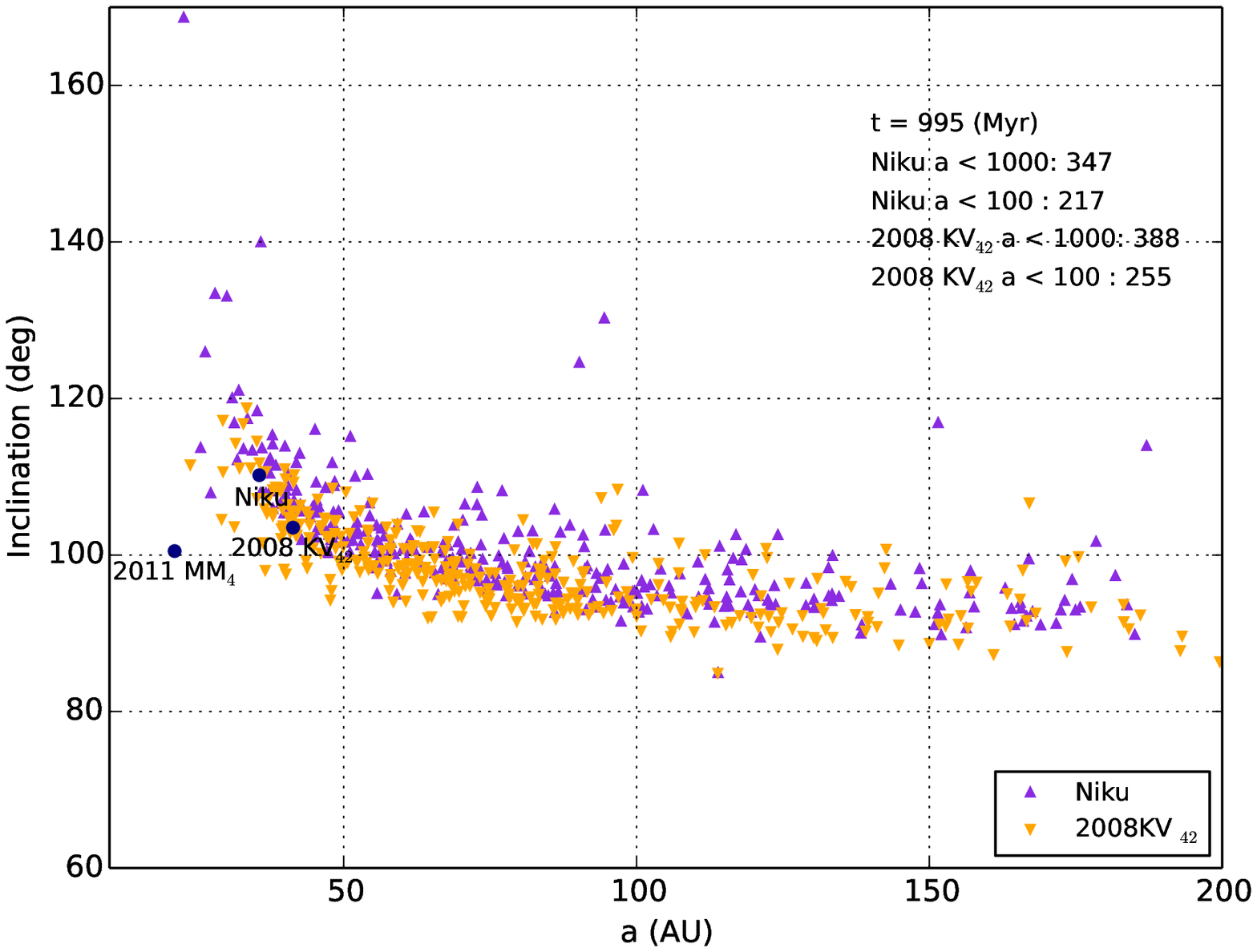}
\caption{The orbital distribution of survivors from 1000 initial clones of Niku (purple) and 2008 KV$_{42}$ (orange) after 1\,Gyr approximately. The overall clones decays far more slowly than normal Centaurs, with $> 30\%$ of the test particles remaining dynamical stability at the end of a 1\,Gyr integration. Note that the orbits of Niku and 2008 KV$_{42}$ are located at the highest density position in this plot. Only a few rare clones could evolve an inclination $<90\arcdeg$. The orbits of selected retrograde objects are also shown.\label{fig:fig1}}
\end{figure}

To explore the evolution of Niku's orbit in the planet-crossing region, we performed
numerical simulations using the {\tt MERCURY} package \citep{Chambers99}. 
We used the covariance matrix generated by the orbit fitting code of \citet{BK00} and generated 1,000 clones drawn from within 3-sigma of the best-fit orbit. 
We included the four giant planets in the simulations, and integrated the 1,000 clones forward for 1\,Gyr using a 180-day time step. 
The majority of the clones were stable for at least 0.1\,Gyr, with the stable half-life being $\sim500$ Myr, with a long tail having lifetimes up to 1\,Gyr. 
The 1,000 clones initially had semi-major axes in the range 35.70\,AU$\, <a<$\,35.74\,AU. 
During the $1$ Gyr simulation, the clones experience orbital evolution, leading to the distribution of \emph{final} semi-major axes illustrated in Figure \ref{fig:fig1}.
Twenty percent of clones with a final orbit $a < 100$ AU stably survive beyond 1\,Gyr; 
thirty percent with $a < 1000$ AU survive beyond 1\,Gyr (see Figure \ref{fig:fig1}). 
This  1\,Gyr lifetime is approximately two orders of magnitude larger than a typical Centaur's lifetime \citep{Volk13}. 

All survivors exhibited similar orbital evolution: 
(1) Their Tisserand parameter with respect to Neptune is similar ($-0.1 <T_{N} < 0.2$), 
(2) most of the clones always have perihelion distances larger than 10\,AU, where they remain beyond the gravitational influence of Saturn and Jupiter, 
(3) the integration of 1,000 clones of 2008 KV$_{42}$, with the same parameters as above, shows a nearly identical result. 

The highest density of survivors in the ($a$, $i$)-plane illustrated in Figure \ref{fig:fig1} matches the location of Niku and 2008 KV$_{42}$. This may hint at the existence of a large population with similar origin. We also checked Niku for the existence of resonances with Neptune, i.e. 5:4 (34.9\,AU) and 4:3 (36.4\,AU): No libration of resonant arguments was observed. 

\begin{figure}
\figurenum{2}
\plotone{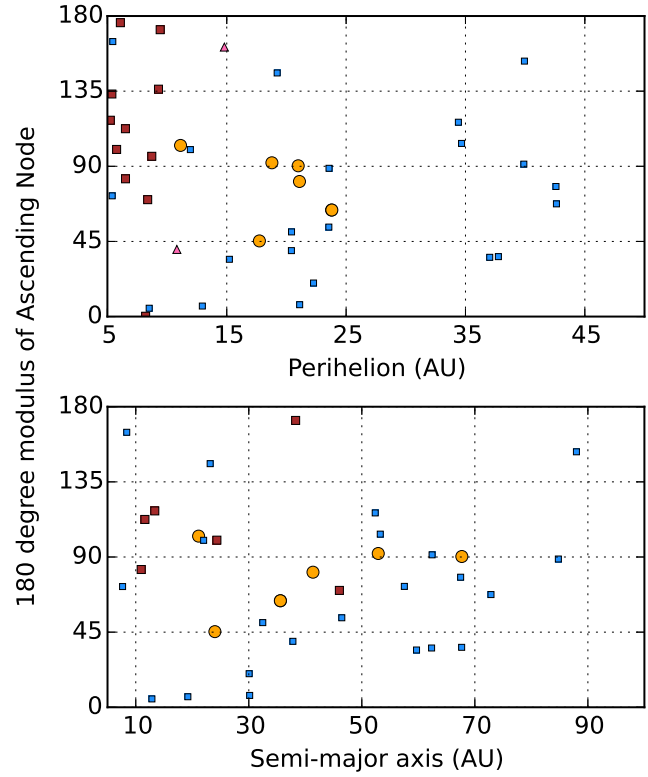}
\caption{The orbits of known Centaurs and scattered-disk objects, whose inclinations more then 2-sigma ($36\arcdeg$) away from the dynamically excited cold classical belt. The blue squares indicate the objects with inclination ($36\arcdeg < i < 60\arcdeg$). The small perihelion objects ($q < 10$) which remain in the gravitational influence of Saturn and Jupiter are shown in red squares. The triangles represent the objects with large semi-major axis ($a > 100$). And the six objects clustered in a common plane are shown by orange circle.\label{fig:fig2}}
\end{figure}

\begin{figure}
\figurenum{3}
\plotone{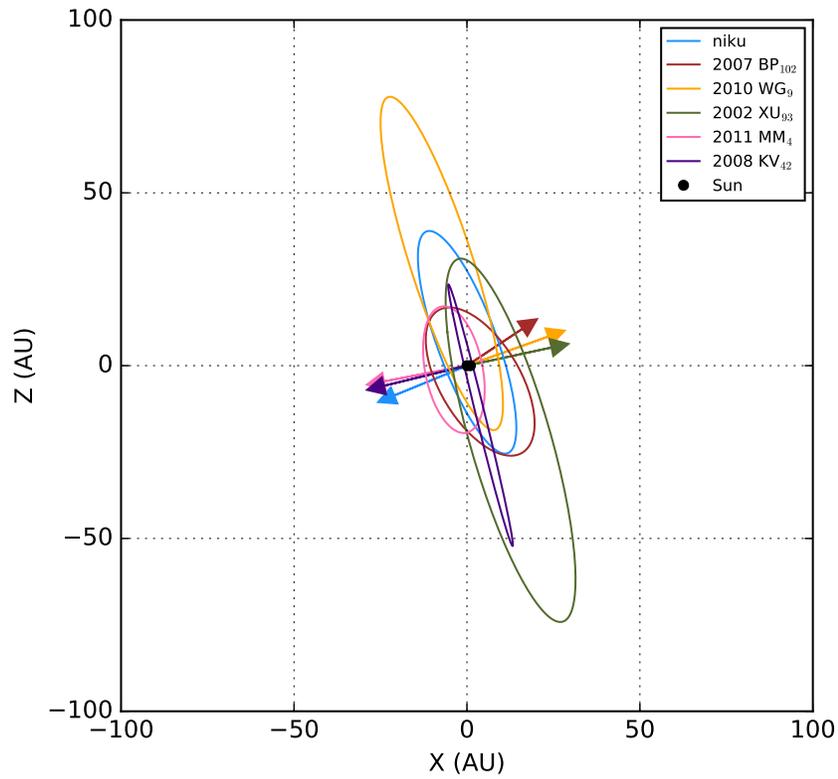}
\caption{The x-z space of selected known objects. The clustering of these six objects in a common plane is obvious.The arrows indicate the directions of the orbital planes. Note that the angular momenta of prograde and retrograde orbits are exactly opposite.\label{fig:fig3}}
\end{figure}

To understand the relation between Niku and other known objects, we performed the following two analyses. 
First, we select known objects from the MPC catalogue, to compare their dynamical evolution with that of Niku. 
Following the criteria ($15<q<30$, $i>70$ and $a<100$) in \citet{Brasser12} and the perihelion evolution we observed in our Niku clones, we use looser constraints with $q > 10$, $a < 100$, $i > 60\arcdeg$ and opposition $\geq 2$ to obtain the sample list for understanding Niku's relation to other similarly inclined objects (Table \ref{tab:tab1}).
If the high inclination reservoir/population mentioned in \citet{Gladman09} and \citet{Brasser12} does indeed exist, then we may find some traces from known objects. For the objects in Table \ref{tab:tab1} with $a < 100$, we observe that there is a clustering in the ascending node ($\Omega$) of the objects, regardless of whether the orbit is prograde or retrograde (see Figure \ref{fig:fig2} and \ref{fig:fig3}). 
The ascending node of the prograde orbits ranges between $45\arcdeg$ and $95\arcdeg$; the ascending node of the retrograde orbit ranges between $243\arcdeg$ and $282\arcdeg$. 
These two ranges are planar opposite, which means the orbits of these six objects occupy an approximately common plane. Note that the angular momenta of the prograde and retrograde orbits are antialigned.
If we change the selection criteria to include all objects with  $q > 5$ and $i > 60\arcdeg$, then we no longer see  any obvious clustering in $\Omega$ (see Figure \ref{fig:fig2}). 

Second, we performed additional numerical simulations for objects in Table \ref{tab:tab1}.
The integrations use the same parameters as described above.
We have verified that the clones of 2010 WG$_{9}$, 2002 XU$_{93}$ and 2008 KV$_{42}$ could all survive within 1000\,AU for 1\,Gyr, with a survival rate of $16\%$, $25\%$ and $38\%$, respectively. 
The clones of objects with smaller $q$ and $a$ (2007 BP$_{102}$ and 2001 MM$_{4}$) have a much smaller probability ($<0.5\%$) of surviving 
until the end of the 1\,Gyr integration.

We note that the precession directions 
of the prograde and retrograde orbits are \emph{opposite}. 
In our integrations, the common plane disappears on 
a short time scale (a few Myr). 

\section{Discussion} \label{sec:discussion}

It is essential to determine the likelihood that the apparent clustering in the longitude of ascending node occurs by chance. Here we discuss possible reasons for, and origins of, this clustering.

\subsection{Coincidence} \label{subs: coin}
Because of the small number of samples (only six members) we can not 
completely reject the hypothesis that the occupation of a common plane is due simply to coincidence. 
Using Monte Carlo simulations, we randomly generate six objects with an isotropic $\Omega$-distribution ($0-360\arcdeg$) and an isotropic $i$-distribution ($0-180\arcdeg$). Then, these objects are separated into two subgroups, 
one with $\Omega\le180\arcdeg$ and one with $\Omega>180\arcdeg$. Finally, two criteria were defined to decide whether they are in a common plane: (1) the $\Omega$ of subgroup with $\Omega < 180\arcdeg$ are within $\pm30\arcdeg$ of average-$\Omega$ ($n$) of this subgroup, and the $\Omega$ of another subgroup ($\Omega > 180\arcdeg$) are within $n + 180 \pm30\arcdeg$, (2) the object with $\Omega < 180\arcdeg$ has a prograde orbit ($i<90\arcdeg$) and the object with $\Omega > 180\arcdeg$ has a retrograde orbit ($i>90\arcdeg$).
In other words, assuming $n$ is the average-$\Omega$ in the range $0\arcdeg<\Omega<180\arcdeg$, then check object if ($n-30\arcdeg < \Omega < n+30\arcdeg~\&~i < 90\arcdeg$) or  ($n+180\arcdeg-30\arcdeg < \Omega < n+180\arcdeg+30\arcdeg~\&~i > 90\arcdeg$). After a million iterations, the probability of getting six objects in a common plane is $0.016\%$, or about 3.8-sigma.
Furthermore, the explanation that the common plane is merely a coincidence becomes even more implausible if we consider 
(a) other orbital parameters and 
(b) the dynamical behavior of the objects. 

Alternatively, one might consider that a distant and inclined primordial disk could be a feeding objects into this high-inclination population. However, any clustering of their ascending nodes would likely be erased within a few Myr by the orbital precession discussed in Section \ref{sec:analysis}. 
At this stage, the small number of known high inclination objects makes it impossible to attempt to reconstruct the orbital distribution of such a putative distant population.




\subsection{Observational Bias} \label{subs: bias}
\begin{figure}
\figurenum{4}
\plotone{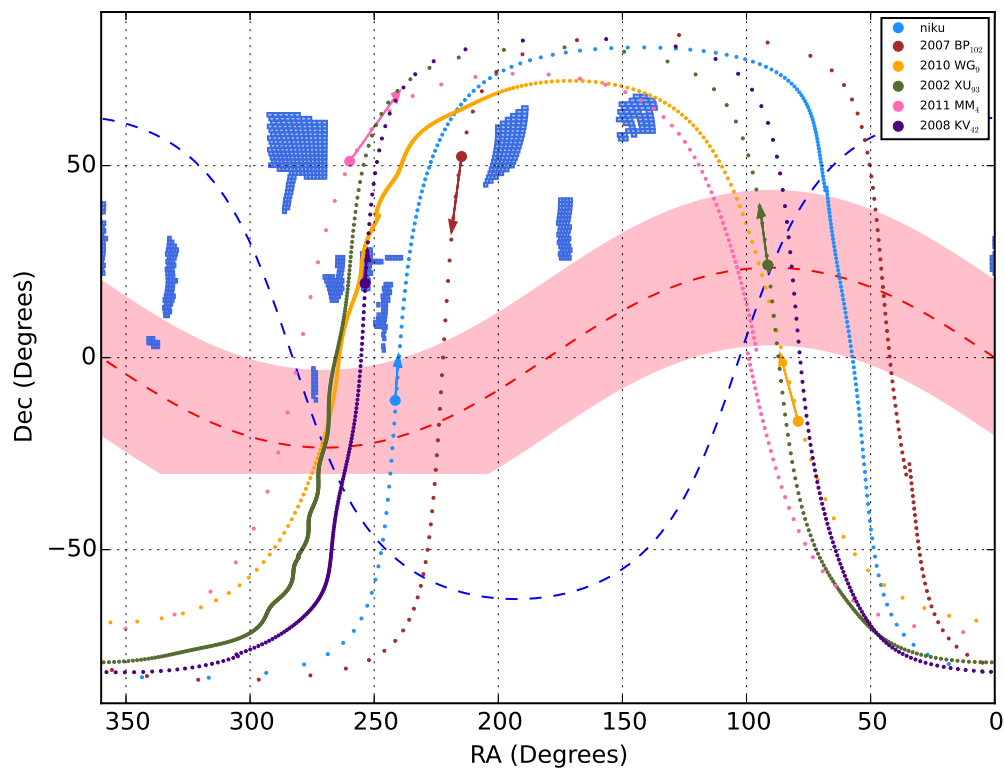}
\caption{The sky path of selected known objects ($q >10$, $a < 100$ and $i > 60$). The arrows indicate the moving directions. The ecliptic plane and galactic plane are showed with dash in red and blue color. The blue patches show the survey region of High Ecliptic Latitude Extension with CFHT FOV. The red belt demonstrates the concept region of PS1 $w_{P1}$-band survey ($\pm20$ ecliptic plane and Dec $> 30\arcdeg$). Note PS1 $w_{P1}$-band survey actually didn't cover the full region of cross section between ecliptic plane and galactic plane. The selected objects show an obvious common plane. \label{fig:fig4}}
\end{figure}

Most of the large surveys which search for moving objects, focus primarily on the region close to the  plane of the ecliptic, typically within $\pm20^{\arcdeg}$. 
The lack of a high ecliptic latitude survey leads to an obvious bias against high inclination objects. 
Except for CFEPS and OSSOS, most of the survey data have not had their detection efficiencies thoroughly characterized, though PS1 will be characterized in the near future. 
A systematic analysis of the expected population is therefore currently impossible. 
The only \emph{high latitude} survey which has been characterized is the CFEPS High Ecliptic Latitude Extension \citep{Kavelaars08,Petit14} (hereafter, CFEPS-HELE). 
The CFEPS-HELE survey detected at least two high inclination objects, namely 2009 MS$_{9}$ and 2008 KV$_{42}$. 
We note that 2009 MS$_{9}$ does \emph{not} occupy the common plane discussed above, but 2008 KV$_{42}$ \emph{does} orbit within this plane. If a survey region only concentrates on a particular RA range, the bias of $\Omega$ and discovered position will be shown in the sky.
Based on the survey regions of PS1 and CFEPS-HELE, we find no obvious bias that can explain the clustering of $\Omega$ (see Figure \ref{fig:fig4}). 

The current version of the pipeline used to generate the catalog of moving object candidates from the PS1 OSS uses a novel heliocentric transformation method to efficiently identify and link observations across multiple nights. This method, to be fully described in Holman \& Payne (2016c \emph{in prep.}), requires a heliocentric distance be assumed for the objects under investigation. The assumed distance is then iterated to find the best \emph{approximate} heliocentric distance, before the linked observations are handed off to have a detailed orbital fit generated by the {\it Orbfit} routine of \citet{BK00}.
Crucially, a \emph{minimum} assumed distance of $25\,AU$ was used in this initial pipeline run, as a means to ensure rapid progress. However it is likely that this biased the initial pipeline search \emph{against} finding Centaurs and other objects with heliocentric distances significantly less than $25\,AU$. As such, we expect that an imminent update to the PS1 OSS, which expand the parameter-space that is searched to well inside $25\,AU$ should reveal additional Centaur-like objects, and may well provide additional high inclination candidates. 
The combination of well characterized PS1 and CFEPS-HELE data sets will provide an extraordinary means to search for a population at high ecliptic latitudes.

\subsection{Planet Nine or Dwarf Planet}  \label{subs: p9}
As discussed in Section \ref{sec:analysis} and Section \ref{subs: coin}, we expect that orbital precession will quickly erase the occupation of a common plane. Hence, another perturber or mechanism may be required to maintain this occupation. Considering the hypotheses from literature, those postulating external forces, like the hypothetical Planet Nine \citep{BB16, Holman16}, Solar companion \citep{Gomes15}, and a dwarf planet in scattered disk \citep{Lykawka08} all seem to be problematic, as they have great difficulty in affecting the planet-crossing region, due to the small perturbations they exert at such great separations (small tidal parameter, $M*AU^{-3}$). 
Our mock integrations using the same parameters in Section \ref{sec:analysis}, but inserting the proposed Planet Nine \citep{BB16}, is \emph{not} able to maintain the common orbital planet of the test particles over any significant time scale, i.e. the orbital precession still erases the $\Omega$ clustering quickly. 

We note that the simulations in \citet{BB16}, provide a source of high-$i$ and large-a TNOs with a clustered distribution of $\Omega$. Their cluster has an ``X'' shape composed of two common planes, and we stress that \emph{neither} of which are coincident with the plane in Figure \ref{fig:fig3}. 
Moreover, even if the objects which occupy our common place did somehow originate from this ``X'' shape, the $\Omega$ clustering would still be expected to vanish due to precession, and hence some mechanism to confine the orbits would be required.

\subsection{Unknown Mechanism or Undetected Dwarf Planet} \label{subs: undet}
The high-$i$ Centaurs and TNOs may originate in the Oort cloud \citep{Brasser12} or some other undetected reservoir \citep{Gladman09}, based on the small change of inclination \citep{Volk13}. 
As mentioned in Section \ref{subs: p9}, irrespective of origin, the observed clustering in  $\Omega$ still requires a mechanism to maintain the common plane in the face of divergent orbital precession. 
The existence of such a mechanism has not been established.


The detailed exploration of such an unknown mechanism is beyond the scope of this Letter, but as indicated in Section~\ref{subs: p9}, we established through numerical integrations that the putative Planet Nine was unable to explain the orbital confinement. 
In addition, we also attempted more extreme alternative scenarios, consisting of integrations that include a synthetic high-inclination dwarf planet of a few Earth masses whose orbit crosses the giant planet region. 
None of these attempts succeeded in anchoring the $\Omega$ of the test particles. Moreover, adding a planet into planet-crossing region has a very high chance of disrupting the orbital structure of the Kuiper Belt and remaining outer Solar System. 

A more detailed set of investigations is required to understand whether this common orbital plane is dynamically robust and long-lived, and if so, what mechanisms contribute to its longevity.
Additionally, it remains to be seen whether the observed common orbital plane survives further observational scrutiny. Well-characterized observations of objects in this plane will also contribute to an understanding of overall observational bias. 
A deeper and wider survey, such as LSST, may provide a means to detect additional high-$i$ objects in this common plane, and/or an undetected dwarf planet that is sculpting their orbits.

\section{Conclusion}  \label{sec:conclusions}
We report the discovery by the Pan-STARRS-1 Outer Solar System Survey of the second retrograde TNO. 
Our numerical integrations of Niku show that it's dynamical evolution is very similar to that of 2008 KV42. 
This result may hint at the existence of a large population of similar origin. 

We have also uncovered the possible occupation of a common plane by the known objects with $q > 10$, $a < 100$ and $i > 60\arcdeg$. The mechanism causing and maintaining this common plane is still unknown. The detection of additional high inclination objects in future surveys, such as PS2 or LSST, will provide additional clues as to the dynamical origin of this population.

%
\acknowledgments
We are grateful to the CFEPS team for kindly providing the pointings of CFEPS-HELE.
We thank all of the staff at the Lulin observatory of the National Central University.
This work was supported by MOST 104-2119-008-024 (TANGO-II) and MOE under the Aim for Top University Program NCU, and Macau Technical Fund: 017/2014/A1 and 039/2013/A2. HWL acknowledges the support of the CAS Fellowship for Taiwan-Youth-Visiting-Scholars under the grant no.2015TW2JB0001.
The Pan-STARRS1 Surveys (PS1) have been made possible through contributions of the Institute for Astronomy, the University of Hawaii, the Pan-STARRS Project Office, the Max-Planck Society and its participating institutes, the Max Planck Institute for Astronomy, Heidelberg and the Max Planck Institute for Extraterrestrial Physics, Garching, The Johns Hopkins University, Durham University, the University of Edinburgh, Queen's University Belfast, the Harvard-Smithsonian Center for Astrophysics, the Las Cumbres Observatory Global Telescope Network Incorporated, the National Central University of Taiwan, the Space Telescope Science Institute, the National Aeronautics and Space Administration under Grant No. NNX08AR22G issued through the Planetary Science Division of the NASA Science Mission Directorate, the National Science Foundation under Grant No. AST-1238877, and the University of Maryland. This research used the facilities of the Canadian Astronomy Data Centre operated by the National Research Council of Canada with the support of the Canadian Space Agency.

\vspace{5mm}
\facilities{PS1, CFHT, Blanco, LOT}
\software{orbfit \citep{BK00} and Mercury 6.2 \citep{Chambers99}}

\clearpage

\clearpage

\begin{deluxetable}{c|cccccccc}[b!]
\tablecaption{The selected high inclination TNOs and Centaurs  \label{tab:tab1}}
\tablecolumns{9}
\tablenum{1}
\tablewidth{0pt}
\tablehead{
\colhead{Object} &
\colhead{$q$} &
\colhead{$a$} &
\colhead{$e$} &
\colhead{$i$} &
\colhead{Node} &
\colhead{Arg of peri} &
\colhead{M} &
\colhead{H} \\
\colhead{} & \colhead{(AU)}  & \colhead{(AU)} & \colhead{} &
\colhead{(degree)} & \colhead{(degree)} & \colhead{(degree)} & \colhead{(degree)} & \colhead{(mag)}
}
\startdata
Niku & 23.81 & 35.725 & 0.334 & 110.3 & 243.8 & 322.6 & 23.8 & 7.4\tablenotemark{a}\\
2008 KV$_{42}$ & 21.11 & 41.347 & 0.49 & 103.5 & 261.0 & 133.3 & 333.1 & 8.9\\
2002 XU$_{93}$ & 21.01 & 67.734 & 0.69 & 77.9 & 90.3 & 28.2 & 4.7 & 8.0\\
2010 WG$_{9}$ & 18.77 & 52.95 & 0.645 & 70.3 & 92.1 & 293.1 & 9.3 & 8.1\\
2007 BP$_{102}$ & 17.73 & 24.0 & 0.261 & 64.7 & 45.3 & 125.3 & 18.6 & 10.6\\
2011 MM$_{4}$ & 11.13 & 21.126 & 0.473 & 100.5 & 282.6 & 6.8 & 41.1 & 9.3\\
\enddata
\tablenotetext{a}{This magnitude is calculated from r mag.}
\tablecomments{The orbital parameters are output by the Minor Planet Center and sorted by perihelion distance. }
\end{deluxetable}

\end{CJK*}

\begin{thebibliography}{}
\bibitem[Alexandersen et al.(2014)]{Alex14} Alexandersen, M., Gladman, B., Kavelaars, J.~J., et al.\ 2014, arXiv:1411.7953 
\bibitem[Bannister et al.(2015)]{Bannister15} Bannister, M.~T., Kavelaars, J.~J., Petit, J.-M., et al.\ 2015, arXiv:1511.02895 
\bibitem[Batygin \& Brown(2016)]{BB16} Batygin, K., \& Brown, M.~E.\ 2016, \aj, 151, 22 
\bibitem[Bernstein \& Khushalani(2000)]{BK00} Bernstein, G., \& Khushalani, B.\ 2000, \aj, 120, 3323 
\bibitem[Brasser et al.(2012)]{Brasser12} Brasser, R., Schwamb, M.~E., Lykawka, P.~S., \& Gomes, R.~S.\ 2012, \mnras, 420, 3396 
\bibitem[Chambers(1999)]{Chambers99} Chambers, J.~E.\ 1999, \mnras, 304, 793 
\bibitem[Gladman et al.(2009)]{Gladman09} Gladman, B., Kavelaars, J., Petit, J.-M., et al.\ 2009, \apjl, 697, L91 
\bibitem[Gomes et al.(2005)]{Gomes05} Gomes, R.~S., Gallardo, T., Fern{\'a}ndez, J.~A., \& Brunini, A.\ 2005, Celestial Mechanics and Dynamical Astronomy, 91, 109 
\bibitem[Gomes et al.(2015)]{Gomes15} Gomes, R.~S., Soares, J.~S., \& Brasser, R.\ 2015, \icarus, 258, 37 
\bibitem[Granvik et al.(2009)]{Granvik09} Granvik, M., Virtanen, J., Oszkiewicz, D., \& Muinonen, K.\ 2009, Meteoritics and Planetary Science, 44, 1853 
\bibitem[Gwyn et al.(2012)]{Gwyn12} Gwyn, S.~D.~J., Hill, N., \& Kavelaars, J.~J.\ 2012, \pasp, 124, 579 
\bibitem[Holman et al.(2015)]{Holman15} Holman, M.~J., Chen, Y.-T., Lin, H.-W., et al.\ 2015, AAS/Division for Planetary Sciences Meeting Abstracts, 47, 211.12 
\bibitem[Holman \& Payne(2016)]{Holman16} Holman, M.~J., \& Payne, M.~J.\ 2016, arXiv:1603.09008 
\bibitem[Kavelaars et al.(2008)]{Kavelaars08} Kavelaars, J.~J., Gladman, B., Petit, J., Parker, J.~W., \& Jones, L.\ 2008, Bulletin of the American Astronomical Society, 40, 47.02 
\bibitem[Levison \& Duncan(1997)]{Levison97} Levison, H.~F., \& Duncan, M.~J.\ 1997, \icarus, 127, 13 
\bibitem[Levison \& Morbidelli(2003)]{Levison03} Levison, H.~F., \& Morbidelli, A.\ 2003, \nat, 426, 419 
\bibitem[Levison et al.(2006)]{Levison06} Levison, H.~F., Duncan, M.~J., Dones, L., \& Gladman, B.~J.\ 2006, \icarus, 184, 619 
\bibitem[Levison et al.(2008)]{Levison08} Levison, H.~F., Morbidelli, A., Van Laerhoven, C., Gomes, R., \& Tsiganis, K.\ 2008, \icarus, 196, 258 
\bibitem[Lykawka \& Mukai(2007)]{Lykawka07} Lykawka, P.~S., \& Mukai, T.\ 2007, \icarus, 189, 213 
\bibitem[Lykawka \& Mukai(2008)]{Lykawka08} Lykawka, P.~S., \& Mukai, T.\ 2008, \aj, 135, 1161 
\bibitem[Magnier et al.(2013)]{Magnier13} Magnier, E.~A., Schlafly, E., Finkbeiner, D., et al.\ 2013, \apjs, 205, 20 
\bibitem[Petit et al.(2014)]{Petit14} Petit, J.-M., Kavelaars, J.~J., Gladman, B., Jones, L., \& Parker, J.\ 2014, AAS/Division for Planetary Sciences Meeting Abstracts, 46, 507.07 
\bibitem[Rabinowitz et al.(2013)]{Rabinowitz13} Rabinowitz, D., Schwamb, M.~E., Hadjiyska, E., Tourtellotte, S., \& Rojo, P.\ 2013, \aj, 146, 17 
\bibitem[Schlafly et al.(2012)]{Schlafly12} Schlafly, E.~F., Finkbeiner, D.~P., Juri{\'c}, M., et al.\ 2012, \apj, 756, 158 
\bibitem[Tiscareno \& Malhotra(2003)]{Tiscareno03} Tiscareno, M.~S., \& Malhotra, R.\ 2003, \aj, 126, 3122
\bibitem[Tonry et al.(2012)]{Tonry12} Tonry, J.~L., Stubbs, C.~W., Lykke, K.~R., et al.\ 2012, \apj, 750, 99 
\bibitem[Volk \& Malhotra(2013)]{Volk13} Volk, K., \& Malhotra, R.\ 2013, \icarus, 224, 66 
\end{thebibliography}
\end{document}